\documentstyle[12pt,overcite,mg7procs,twoside]{article}

 \def\tenrm{\small}
\def\mathput#1{\relax \ifmmode \displaystyle #1\else $\displaystyle #1$\fi}

\makeatletter \long\def\ylap#1{\vbox to \z@{\vss#1\vss}} \makeatother     

\def\Vtop#1{\vtop{\hsize=3in \baselineskip=15pt 
                 \def\\{\hfil\break}\noindent\hangindent 20pt\strut #1\strut}}

\def\VVVtop#1{\vtop{\hsize=1.75in \baselineskip=15pt
                              \def\\{\hfil\break}\noindent #1}}

\input prepictex \input pictex \input postpictex

\def\half{\frac12}  \def\sub#1{_{\hbox{$#1$}}}

  \def\del{\nabla}               
    \def\curl{\mathop{\rm curl}\nolimits}

\catcode`@=11     % make @ like a letter
     
\def\eqalign#1{\null\,\vcenter{\openup\jot\m@th\def\\{\crcr}
  \ialign{\strut\hfil$\displaystyle{##}$&$\displaystyle{{}##}$\hfil
      \crcr#1\crcr}}\,}

\newcommand{\pub}{  \setbox43=\vbox{
   \flushleft{
  {\footnotesize Proc.\ Seventh Marcel Grossmann Meeting (1994), 
        Eds. R.T Jantzen and G.M. Keiser}\\[-2pt] 
  {\footnotesize \copyright 1996 World Scientific, pp.~133--152 [reformatted 2001]}\\
	 }}
  \vbox to 0pt{\vskip-22.5pt\vskip-35pt 
    \hbox to \textwidth{\vbox to 8.5pt{}\box43}\vss}
  \nointerlineskip} 

\begin{document}
\pub

\begin{Titlepage}
\Title
GRAVITOELECTROMAGNETISM: JUST A BIG WORD?
\endTitle
\Author{ROBERT T. JANTZEN}
Department of Mathematical Sciences, Villanova University, 
Villanova, PA 19085, USA, and\\
ICRA, Dipartimento di Fisica, University of Rome, I--00185 Roma, ITALY
\endAuthor
\Author{PAOLO CARINI}
  GP-B, Hansen Labs, Stanford University, Stanford, CA 94305, USA and\\
  International Center for Relativistic Astrophysics, %\\
  University of Rome, I--00185 Roma, Italy
\endAuthor
\And
\Author{DONATO BINI}
  Physics Department and
  International Center for Relativistic Astrophysics, %\\
  University of Rome, I--00185 Roma, Italy
\endAuthor
\endAuthors

\begin{Abstract}
Arguments are made in favor of broadening the scope of the various
approaches to splitting spacetime into a single common framework in which
measured quantities, derivative operations, and adapted
coordinate systems are clearly understood
in terms of associated test observer families.
This ``relativity of splitting formalisms" 
for fully nonlinear gravitational theory
has been tagged with the
name ``gravitoelectromagnetism" because of the well known 
analogy between its linearization and electromagnetism,
and it allows relationships between
the various approaches to be better understood and makes it easier
to extrapolate familiarity with one approach to the others. This is
important since particular problems or particular features of those
problems in gravitational theory
are better suited to different approaches, and the present
barriers between the proponents of each individual
approach sometimes prevent the best match from occurring.
\end{Abstract}

\end{Titlepage}

%%%%%%%%%%%%%%%%%%%%%%%%%%%%%%%%%%%%%%%%%%%%%%%%%%%%%%%%%%%%%%%%%%%%%%%%%
\section{Introduction}

Before I explain what I mean by this big word ``gravitoelectromagnetism"
which is still not
a part of our relativity jargon, I must admit my reluctance to give a
plenary talk on mathematical formalism, which is what I am about to do. 
I am not specialized in the various difficult and challenging
physical problems that my remarks may touch upon, 
but I believe that the
mathematical tools of GEM do help us better understand the way in which
spacetime structure enters some of those difficult calculations. 
To justify my talk I can look to the short objective statement found at
the beginning of the proceedings of each Marcel Grossmann meeting, 
where one finds the two phrases
``$\ldots$ emphasis on mathematical foundations $\ldots$"    
and
``$\ldots$ deepen our understanding of spacetime structure $\ldots$".
The conference itself derives its name from the standard lore about 
Einstein's mathematician friend
who supposedly pointed out some mathematical
tools crucial for the development of general relativity.

In a similar way I wish to point out some mathematical tools for splitting
spacetime which together are important in many different applications in
general relativity and which separately are continuously being used, 
but which have failed to find a common home as a standard part of the 
foundations of general relativity
and are not even always recognized when they are being used.
%that one finds say, in textbooks and graduate courses.

There are two difficulties which limit the accessibility of this whole set
of tools.

\begin{itemize}
\item The first is the existence of a ``literature horizon" beyond 
which many of the books and articles which do discuss these tools have 
fallen. 
The quantity of literature in relativity and gravitation, as in other fields,
has wildly grown over the decades and many results are simply buried in the
past, lost in the sheer volume of publications. 
References familiar to
one generation are often less so or unknown to the successive one, an
effect which increases with the passage of time.

\item The second difficulty is, even if one is successful in 
breaking through the literature horizon and in mining its hidden wealth 
for the relevant gems, 
one often finds either antiquated notation or lack of a common
mathematical framework in which to formulate a coherent 
approach to all of the 
various pieces one finds, some familiar and some not.
\end{itemize}

What we have attempted to do is to describe such a common framework and
notation, making more precise certain notions 
and their relationships to each other,  
and it seems natural to use the word ``gravitoelectromagnetism" (GEM)
to refer to this way of looking at general relativity.
With the increasing widespread use of the terms ``gravitoelectric (GE) field",
``gravitomagnetic (GM) field", and ``gravitomagnetism", 
I think most of us are
aware of the analogy between the linear theory of electromagnetism 
and
the linearized theory of general relativity \cite{for}
and related approximate theories, 
an analogy which seems to lend itself immediately the longer word
``gravitoelectromagnetism." 
The analogy is in fact with the space-plus-time splitting of electromagnetism
in flat spacetime. 
The GE field is associated with local accelerations
and in the post-Newtonian limit, related to the Newtonian gravitational field, 
while the GM field is associated with local
rotations and the so called effect of the ``dragging of inertial frames,"
and is a new feature of relativistic gravitational theories compared to
Newtonian gravity. 

%%%%%%%%%%%%%%%%%%%%%%%%%%%%%%%%%%%%%%%%%%%%%%%%%%%%%%%%%%%%%%%%%%%%%%%%%%%%%
% figure 1:
\begin{figure}
\font\bobbig= cmr17 scaled \magstep1
 \setbox1=\hbox{\bobbig E} %\showthe\ht1  % 20.4 pt
$$
\hbox{\bobbig G R} \quad \Rightarrow \quad
\vcenter{\offinterlineskip
\halign{ #\hfil & #\hfil & #\hfil & #\hfil\cr
spatial  & {\bobbig G}eometry    &
  $\left. \vrule height .7\ht1 depth 0pt width 0pt \right\}$ 
   & observer-space geometry \cr
gravito- & {\bobbig E}lectricity & & \cr \noalign{\vskip 1pt}
         &                    & 
  \ylap{\hbox{$\left.\vrule height .55\ht1 depth .45\ht1 width0pt \right\} $}}
  & \ylap{\hbox{observer kinematics}} \cr \noalign{\vskip 1pt}
gravito- & {\bobbig M}agnetism   & \cr
}
}
$$
\caption{GEM: Spacetime Splitting as a Nonlinear Analogy/Generalization of 
Electromagnetism, or the Relativity of Splitting Formalisms}
\end{figure}
%%%%%%%%%%%%%%%%%%%%%%%%%%%%%%%%%%%%%%%%%%%%%%%%%%%%%%%%%%%%%%%%%%%%%%%%%%%%%

The splitting of fully nonlinear general relativity gives rise to a
nonlinear analogy with flat space electromagnetism 
whose linearization
leads to the more familiar linear analogy, but this is not well known at all. 
The rich structure of general relativity and curved spacetime allows many
variations of the simple act of splitting spacetime into space plus time. 
Unfortunately these variations have developed in isolation from each
other in styles which make it difficult to relate to each other or analyze
in terms of their geometric relationships.
Each of the various approaches to ``splitting spacetime" is equivalent to
describing what a family of test observers in spacetime measure along
a certain family of trajectories in spacetime, with respect to a certain class
of adapted spacetime frames evolving in a specified way along those 
trajectories. These latter assumptions are equivalent to the choice of
derivative used to measure evolution along that  family of trajectories.

Of course to treat these questions properly, one needs to introduce
a precise mathematical description which necessarily involves some
investment of time to become familiar with,
but the end result is a concise framework within which one can 
unambiguously study otherwise confusing issues.
The details are completely straightforward. One need only reformulate
the mathematics of special relativity, usually treated in terms of the
affine structure of Minkowski spacetime,
in such a way that it respects the manifold structure of this spacetime,
leading automatically to the appropriate (often multiple)
generalizations to curved spacetimes.
This natural  marriage of special relativity, curved manifolds,
and modern mathematical methods, and the interpretation of adapted
coordinate systems in this context,
though not conceptually difficult, has not found its way into the
standard toolbox of relativists, even though parts of it find
widespread application in gravitational physics.

Before discussing the possibilities in more detail, 
it is useful to mention
three collections of  names which are associated with the most visible
splitting approaches, 
each of which is anchored in the literature by high profile books or articles.

\begin{enumerate}
\item
ADM: Arnowit-Deser-Misner\cite{adm},
MTW {\it Gravitation\/}\cite{mtw} (Wheeler\cite{whe}: lapse and shift). 
%\hfil\break
[two of the most well known anagrams in relativity] 
\hfil\break
[ADM approach motivated by quantum gravity]
%cosmology, quantum gravity, numerical relativity, black holes] 
\hfil\break
``slicing approach"

\item
Landau-Lifshitz, {\it Classical Theory of Fields}\cite{ll}. 
\hfil\break
[roots in 40's edition, reports stationary case of 50's work] 
\hfil\break
``threading approach"

\item
Ehlers-Hawking-ELLIS, cosmology review articles\cite{ehl,haw,ell71,ell73}. 
\hfil\break
[kinematical quantities of a unit timelike vector field] 
\hfil\break
``congruence approach"
\end{enumerate}

Of course these are only the tip of the iceberg so to speak, 
with foundations
whose ``first generation" of authors might also include among many others: 
Einstein, Bergmann, Lichnerowicz, M\"oller, Zel'manov,
Cattaneo, Ferrarese, Choquet-Bruhat, and Dirac (see Ref.~[\citen{mfg}]
for references).

Each of these three approaches ignores the existence of the others. 
Each has its own peculiar established notation that makes comparisons more
difficult. 
Each has certain applications which seem more natural arenas for their use.

Listing some ongoing applications helps justify
giving some attention here to the
whole idea of spacetime splittings and their relationships. 
In these applications and many others one or more
of these splittings naturally
occurs or is an important tool. 
Though Einstein made a great leap forward by unifying space and time into
a single object, 
we can only experience it 
through our space-plus-time perspective as observers within it, 
and such splittings help us interpret spacetime geometry in terms of that
perspective. 
This is probably the most important reason why spacetime
splittings occur so frequently in general relativity.
Table 1 is a short list of some topics in which spacetime splittings play
an important role.

%%%%%%%%%%%%%%%%%%%%%%%%%%%%%%%%%%%%%%%%%%%%%%%%%%%%%%%%%%%%%%%%%%%%%%%%%%%%%
\begin{table}
\vbox{%
\begin{tabular}{|l|l|l|}\hline\strut
classical GR foundations & 
  \Vtop{initial value problem, degrees of freedom, dynamics} & 1\\ \hline\strut
quantum gravity & canonical approach, Ashtekar variables & 1\\ \hline\strut
minisuperspace cosmology & 
  \Vtop{exact solutions, qualitative analysis, multidimensional theories,
        classical and quantum} & 1\\ \hline\strut
exact solution techniques & 2 or 1 Killing vector cases & 2, 1\\ \hline\strut
black holes & membrane paradigm, numerical work & 1\\ \hline\strut
perturbation problems: & & \\ \hline\strut
\quad FRW gauge invariant & Bardeen & 1 \\ \strut
\qquad cosmology & Ellis, Bruni, et al & 3 \\ \hline \strut
  {\quad isolated systems} & \Vtop{Ehlers et al, Newtonian limit, PN initial
         value problem} & 2, 3 \\ \hline\strut
   & \Vtop{Damour et al\\ PN Celestial Mechanics} & 2\\ \hline\strut
   & \Vtop{Thorne, Forward, PN Theory} & 2\\ \hline\strut
inertial forces & Abramowicz et al & 2\\ \hline\strut
parametric manifolds & Perjes, Boersman and Dray & 2\\ \hline
\end{tabular}
}
\caption{Some applications of spacetime splitting. The number in the last
column refers to the three splitting schools listed in the text.}
\end{table}
%%%%%%%%%%%%%%%%%%%%%%%%%%%%%%%%%%%%%%%%%%%%%%%%%%%%%%%%%%%%%%%%%%%%%%%%%%%%%

Okay, so perhaps we can agree that spacetime splitting is a widespread
activity, 
although those of you who actually do it probably think whatever you are
presently doing is just fine 
and we don't really need to talk about the bigger picture. 
It is exactly this attitude which has maintained the fragmentation
that has characterized the topic for decades. 
It would be fun to try to trace the history of these ideas in detail, 
but I think it is a better investment of time to try  and communicate
some sense of what a common framework for them is 
and how it can help us better understand certain aspects of general relativity.

%%%%%%%%%%%%%%%%%%%%%%%%%%%%%%%%%%%%%%%%%%%%%%%%%%%%%%%%%%%%%%%%%%%%%%%%%%%%%
\section{Splitting: The Basics}

First, given the unified concept of spacetime which is the arena of general
relativity, 
what does it mean to speak of space and time separately? 
These are in fact complementary notions related to two distinct ways of
characterizing time itself. 
The first of these is embodied in our wristwatches that most of us are
probably wearing. 
It is our own local time reference that we carry with
us whereever we go and use to mark off events along our worldline in
spacetime. 
The second notion might be exemplified by a VCR setting to record a TV
program (an analogy not meant to be taken too seriously!).
At a certain moment of time within a given geographical area, the
program begins, 
for everyone in that area that cares to tune in. 
This is a synchronization of their local times, which is also a way of
identifying the concept of space within spacetime.

In order to get started, let's use these two notions of time to give
a broad characterization of the different splitting approaches,
as sketched in Table 2.
One can either do a partial splitting or a full splitting.
A partial splitting (choice of time gauge) is based on
a choice of independent local time or 
globally synchronized time (space)
respectively made by specifying a distribution
of local time directions through a unit timelike vector field $u$, or
by specifying an integrable distribution of spacelike local rest spaces
$LRS_u$, where $u$ is a vorticity-free unit timelike vector field orthogonal
to each such space. 
Both partial splittings are equivalent to specifying only a timelike
congruence (the worldlines of $u$) or only a spacelike slicing
(the integrable hypersurfaces of the distribution of 3-spaces orthogonal
to $u$).
The additional complementary choice of space in a full splitting
determines the spatial gauge freedom for the given choice of time
in each category, so that one has a pair consisting of a slicing and
a threading with the causality condition imposed on the one associated
with the choice of time.
A full splitting is most easily described locally by using an adapted
coordinate system $\{t,x^a\}$ which incorporates the additional structure of
a choice of parametrization for
the family of slices (time function), 
and a choice of parametrization for the threading congruence curves
(spatial coordinate system). It seems reasonable to call this structure
modulo spatial coordinate transformations a parametrized nonlinear
reference frame, where the word ``parametrization" refers to the
specific choice of time function.
Some of the variations of the main splitting schemes found in the
older literature depend on this additional time function rather
than just the family of slices.

%%%%%%%%%%%%%%%%%%%%%%%%%%%%%%%%%%%%%%%%%%%%%%%%%%%%%%%%%%%%%%%%%%%%%%%%%%%
\begin{table}
\typeout{4 pictex pictures for table 2}
\def\Unit{0.5cm}
\typeout{solid threads}
\setbox43=\vbox to 0pt{
\beginpicture  \setcoordinatesystem units <\Unit,\Unit> point at 0 0   
  \setquadratic

  \plot 0 0      .2 2  1 4 /  
  \plot 1 -0.2  1.2 2  2 4 /  
  \plot 2 0     2.2 2  3 4 /  

  \setsolid \setlinear

  \arrow <.3cm> [.15,.4]    from  1 4 to 1.15 4.3 
  \arrow <.3cm> [.15,.4]    from  2 4 to 2.15 4.3 
  \arrow <.3cm> [.15,.4]    from  3 4 to 3.15 4.3 
      
\endpicture}

\setbox44=\vbox{
\beginpicture  \setcoordinatesystem units <\Unit,\Unit> point at 0 0   
  \setquadratic

% bot
  \plot -0.7 1     .4  .9  1.5  .5 / 
  \plot  1.5 .5   2.6 1.1  3.8 1.2 /
  \plot -0.7 1.0 -0.1 1.35 .6 1.57 /
  \plot  2.3 1.6 3.1 1.45 3.8 1.2 /
% mid 
  \plot -0.6 1.7   .5 1.6  1.6 1.2 /
  \plot  1.6 1.2  2.7 1.8  3.9 1.9 /
  \plot -0.6 1.7 0.0 2.05 .7 2.27 /
  \plot  2.4 2.3 3.2 2.15 3.9 1.9 /

% top: bot L R top L R   
  \plot -0.5 2.4   .6 2.3  1.7 1.9 /
  \plot  1.7 1.9  2.8 2.5  4.0 2.6 /
  \plot -0.5 2.4  0.5 3.0  1.4 3.3 /
  \plot  1.4 3.3  2.7 3.0  4.0 2.6 /
  
  \setdashes
% threads    
  \plot 0 0      .2 2  1 4 /  
  \plot 1 -0.2  1.2 2  2 4 /  
  \plot 2 0     2.2 2  3 4 /  
% arrows
  \setsolid \setlinear
  \arrow <.3cm> [.15,.4]    from  1 4 to 1.15 4.3 
  \arrow <.3cm> [.15,.4]    from  2 4 to 2.15 4.3 
  \arrow <.3cm> [.15,.4]    from  3 4 to 3.15 4.3     
\endpicture}

\setbox45= \vbox{
\beginpicture  \setcoordinatesystem units <\Unit,\Unit> point at 0 0   
  \setquadratic 
  \setdashes

% bot
  \plot -0.7 1     .4  .9  1.5  .5 / 
  \plot  1.5 .5   2.6 1.1  3.8 1.2 /
  \plot -0.7 1.0 -0.1 1.35 .6 1.57 /
  \plot  2.3 1.6 3.1 1.45 3.8 1.2 /
% mid 
  \plot -0.6 1.7   .5 1.6  1.6 1.2 /
  \plot  1.6 1.2  2.7 1.8  3.9 1.9 /
  \plot -0.6 1.7 0.0 2.05 .7 2.27 /
  \plot  2.4 2.3 3.2 2.15 3.9 1.9 /

% top: bot L R top L R   
  \plot -0.5 2.4   .6 2.3  1.7 1.9 /
  \plot  1.7 1.9  2.8 2.5  4.0 2.6 /
  \plot -0.5 2.4  0.5 3.0  1.4 3.3 /
  \plot  1.4 3.3  2.7 3.0  4.0 2.6 /
  
\setsolid
% threads    
  \plot 0 0      .2 2  1 4 /  
  \plot 1 -0.2  1.2 2  2 4 /  
  \plot 2 0     2.2 2  3 4 /  
% arrows
  \setsolid \setlinear
  \arrow <.3cm> [.15,.4]    from  1 4 to 1.15 4.3 
  \arrow <.3cm> [.15,.4]    from  2 4 to 2.15 4.3 
  \arrow <.3cm> [.15,.4]    from  3 4 to 3.15 4.3     
\endpicture}

\typeout{solid slices}
\setbox46= \vbox{
\beginpicture  \setcoordinatesystem units <\Unit,\Unit> point at 0 0   
  \setquadratic

% bot
  \plot -0.7 1     .4  .9  1.5  .5 / 
  \plot  1.5 .5   2.6 1.1  3.8 1.2 /
  \plot -0.7 1.0 -0.1 1.35 .6 1.57 /
  \plot  2.3 1.6 3.1 1.45 3.8 1.2 /
% mid 
  \plot -0.6 1.7   .5 1.6  1.6 1.2 /
  \plot  1.6 1.2  2.7 1.8  3.9 1.9 /
  \plot -0.6 1.7 0.0 2.05 .7 2.27 /
  \plot  2.4 2.3 3.2 2.15 3.9 1.9 /

% top: bot L R top L R   
  \plot -0.5 2.4   .6 2.3  1.7 1.9 /
  \plot  1.7 1.9  2.8 2.5  4.0 2.6 /
  \plot -0.5 2.4  0.5 3.0  1.4 3.3 /
  \plot  1.4 3.3  2.7 3.0  4.0 2.6 /
      
\endpicture}

%   vertically centered vbox:
\long\def\vc#1{\vbox to 1.5truein{\hsize= 1.7truein\parindent=0pt\vfil
 \strut #1\strut\vfil}}
\long\def\vcV#1{\vbox to .7truein{\hsize= 1.7truein\parindent=0pt\vfil
 \strut #1\strut\vfil}}
\long\def\vcs#1{\vbox to .7truein{\hsize= 1.5truein\parindent=0pt\vfil
 \strut #1\strut\vfil}}
% modification of 12pt strutbox of plain.tex:
  \newbox\strutboxfifteen
  \setbox\strutboxfifteen=\hbox{\vrule height 13pt depth 8pt width0pt}
\def\strutfifteen{\relax \ifmmode\copy\strutboxfifteen
\else \unhcopy\strutboxfifteen\fi}
 \let\strutf=\strutfifteen
 \let\strutF=\strut
  \newbox\strutboxhigh
  \setbox\strutboxhigh=\hbox{\vrule height 20pt depth 8pt width0pt}
\def\struthigh{\relax \ifmmode\copy\strutboxhigh
\else \unhcopy\strutboxhigh\fi}
 \let\struth=\struthigh
\setbox0=\ylap{\vbox{\halign{\strut\hfil # \hfil\cr
                    without\cr causality\cr condition\cr}}}
\ht0=0pt \dp0=0pt
$$
\vbox{\tabskip=0pt \offinterlineskip
\def\trule{\noalign{\hrule}}
\halign{\strutf#& \vrule#\tabskip=1em plus2em&
  \hbox to 1.60truein{\hfil #\hfil}& \vrule#&
  \hbox to 1.7truein{\hfil #\hfil}& \vrule#&
  \hbox to 1.75truein{\hfil #\hfil}& \vrule#\tabskip=0pt\cr
\multispan3 &\multispan5\hrulefill\cr
& \multispan2 & & \VVVtop{time: 1\strutf\\(``single-observer" time)\strutf }& & 
                  \VVVtop{space: 3\strutf\\(``moment of time")\strutf} 
                                                                   & \cr\trule
& & \vc{\hsize=1.5truein\halign{\strut #\hfil\cr
          PARTIAL SPLITTING\cr
          $u$ or $LRS_u$\cr  GE, GM fields\cr}} & & 
  \vc{\struth time: 1\par \vfill 
 \ \hskip45pt\raise-55pt\box43 
\vfill  (3) congruence p.o.v.} & & 
  \vc{\struth space: 3\par \vfill
 \ \hskip45pt\raise-55pt\box46 
\vfill
  (4) hypersurface p.o.v.}      &  \cr\trule
& & \vc{\hsize=1.5truein\halign{\strut #\hfil\cr
          FULL SPLITTING\cr
          parametrized nonlinear\cr
          reference frame: $\{t,x^a\}$\cr
          GE, GM fields \cr
          \quad  and potentials\cr}} & & 
  \vc{\hsize=1.65in\struth 
        time $+$ space: 1 $+$ 3\par \vfill 
 \ \hskip45pt\raise-55pt\box45 
\vfill
 (2) threading p.o.v.} & & 
  \vc{\struth space $+$ time: 3 $+$ 1\par \vfill
 \ \hskip45pt\raise-55pt\box44 
\vfill
  (1) slicing p.o.v.}  
                                                                  &  \cr\trule
& & \vcs{TIME gauge} & & 
     \vcV{\VVVtop{timelike observers\\ $\leftrightarrow$ threading\strutf}} & &
     \vcV{\VVVtop{spacelike local rest\strutF\\spaces
             (timelike normal observers)\strutF\\
              $\leftrightarrow$ slicing\strutF}}                   & \cr\trule
& & \vcs{SPACE gauge} & & 
     \vcV{\VVVtop{\strutF arbitrary synchronization\\ of observer times\strutF\\ 
            $\leftrightarrow$ slicing\strutF}} & &
     \vcV{\VVVtop{\strutF arbitrary identification\\ of 
                        ``points of space"\strutF\\
              $\leftrightarrow$ threading\strutF}}             & \cr\trule
}}$$

\caption{A characterization of the different points of view (p.o.v.)
that may be
adopted in splitting spacetime.  Solid lines in diagrams imply the use
of the appropriate causality condition while dashed lines indicate
that no causality condition is assumed. The hypersurface p.o.v.\
is essentially equivalent to the vorticity-free congruence p.o.v.}
\end{table}
%%%%%%%%%%%%%%%%%%%%%%%%%%%%%%%%%%%%%%%%%%%%%%%%%%%%%%%%%%%%%%%%%%%%%%%%%%%%%%%

Given a partial or full splitting of spacetime,
two key ideas characterize the splitting process:
measurement and evolution.
One interprets $u$ as the 4-velocity of a family of test observers which
``measure" both spacetime tensor fields and spacetime differential
operators and tensor equations involving these quantities.
This is done as in special relativity but independently on each
tangent space rather than globally on all of spacetime at once as on
Minkowski spacetime referred to a family of inertial observers in
special relativity theory. It is accomplished simply
by orthogonal projection of everything in sight, based on the
underlying orthogonal decomposition of each tangent space into a local
time direction (along $u$) and an orthogonal local rest space $LRS_u$.
The temporal projection, being associated with a 1-dimensional subspace, 
may be simplified to a
scalar projection, thus leading to a family of ``spatial"
tensor fields or ``spatial" tensor 
operators of different ranks
when decomposing a single spacetime tensor or tensor operator.
The contraction of any index of a spatial tensor field with $u^\alpha$
or $u_\alpha$ is zero.

Evolution is the description of how fields ``evolve in time" and is
equivalent differentially to specifying a direction of
differentiation along which the evolution will take place, easily
given as the tangent to a congruence of evolution curves, as well as
a way of evolving a reference spatial frame along those curves
to measure the evolution against. This information can be packaged
in a single temporal derivative operator along the evolution congruence
which acts on the spaces of spatial fields.

The congruence of measurement worldlines and the congruence of evolution
curves may coincide (congruence, hypersurface, and threading points of view) 
or be independent (slicing point of view). The threading point of view
is just a more explicit representation of the congruence point of view
which takes advantage of the additional (arbitrary) synchronization
information represented by the slicing. The slicing point of view is
instead a 2-congruence approach, one for measurement and one for evolution,
so that the test observers are in relative motion with respect to the
curves describing the evolution. 

One may introduce the quotient space of
the spacetime by the threading curves, or ``computational 3-space,"
and the quotient space by the observer worldlines, or ``observer 3-space."
These coincide except for the slicing point of view. 
In the full
splitting one has a 1-parameter family of embeddings of the computational
3-space into the original spacetime generating the nonlinear reference 
frame and natural
isomorphisms between the computational 3-space tangent spaces 
and the corresponding local rest
spaces associated with the test observers on the spacetime, 
enabling one to consider the spatial measured tensor fields as time-dependent
tensor fields on the computational 3-space.

\section{Splitting: A Few Details}

The measurement process is just an orthogonal decomposition based on
the following representation of the identity tensor in terms of the
temporal and spatial projections associated with the test observers
with 4-velocity $u^\alpha$
\beq
   \delta^\alpha{}_\beta = T(u)^\alpha{}_\beta + P(u)^\alpha{}_\beta 
     = [ - u^\alpha u_\beta ] + [\delta^\alpha{}_\beta + u^\alpha u_\beta ] 
      \ .
\eeq
Applied to a vector field it leads to a scalar and a spatial vector, once
one discards factors of $u^\alpha$ ( or in general of $u_\alpha$ as well)
with free indices
\beq
    X^\alpha = [T(u)X]^\alpha + [P(u)X]^\alpha 
             \leftrightarrow ( - u_\beta X^\beta, P(u)^\alpha{}_\beta X^\beta)
    \ .
\eeq
This decomposition may be extended to any rank tensor field in an obvious
way, yielding a family of spatial tensor fields of all ranks up to the
original rank.

For a single nonzero mass test particle world line with timelike unit
4-velocity $U^\alpha$ and corresponding 4-momentum $P^\alpha = m U^\alpha$, 
this process yields the usual special relativistic
quantities, namely the gamma factor, relative velocity, energy, and spatial
momentum
\beq\eqalign{
  U^\alpha & \leftrightarrow (\gamma(U,u), \gamma(U,u) \nu(U,u)^\alpha)\ , \cr
  P^\alpha & \leftrightarrow (E(U,u), p(U,u)^\alpha) \ , \cr}
\eeq
where $\gamma(U,u) = [1 - ||\nu(U,u)||^2]^{-1/2}$
and $||X|| = |X_\beta X^\beta|^{1/2}$.

For the spacetime covariant and contravariant metrics, the only
nontrivial fields this yields are the spatial such metrics
(let $X^\flat$ and $X^\sharp$ be index-free notation
kernel symbols for tensors whose
indices have all been lowered and raised respectively)
\beq
    P(u)_{\alpha\beta}  = [P(u)g]_{\alpha\beta} \ ,\qquad
    P(u)^{\alpha\beta}  = [P(u)g^\sharp]^{\alpha\beta} \ ,
\eeq
while  the only nonzero field resulting from measuring the 
oriented unit volume 4-form 
$\eta_{\alpha\beta\gamma\delta} = |g|^{1/2}\epsilon_{\alpha\beta\gamma\delta}$
is the unit spatial volume 3-form
\beq
   \eta(u)_{\alpha\beta\gamma} = u^\delta \eta_{\delta\alpha\beta\gamma}
\eeq
which may be used to introduce the spatial cross product of two vectors
\beq
     [X \times_u Y]^\alpha 
         = \eta(u)^\alpha{}_{\beta\gamma} X^\beta Y^\gamma\ .
\eeq
and to introduce the spatial duality operation 
$\ast_u$ on antisymmetric spatial tensor fields.
The spatial dot product is defined analogously
\beq
     X \cdot_u Y  = P(u)_{\alpha\beta} X^\alpha Y^\beta \ .
\eeq

One may also measure differential operators: the covariant derivative
$\nabla$, the exterior derivative $d$, and the Lie derivative $\pounds$.
In this process certain spatial differential operators arise, where a
spatial operator is one which maps the space of spatial tensor fields into
itself. These may be distinguished as spatial or temporal derivative
operators, according to the direction along which they differentiate.
It is convenient to introduce
the spatial covariant derivative $\nabla = P(u)\nabla$, which is a spatial
derivative operator,
and the spatial Lie derivative 
$\pounds(u)_{\hbox{$X$}} = P(u)\pounds_{\hbox{$X$}}$,
from which both Lie derivatives along spatial and temporal directions
may be obtained. In each case all free indices are spatially projected
after the spacetime derivative acts on a tensor field.
The spatial Fermi-Walker derivative along $u$,
$\nabla_{\rm(fw)}(u) = P(u)\nabla_{\hbox{$u$}}$
(so named since it coincides with the spacetime Fermi-Walker derivative
along $u$ when acting on spatial tensor fields),
and the temporal Lie derivative
$\nabla_{\rm(lie)}(u) = P(u)\pounds_{\hbox{$u$}}$
are both temporal derivative operators.
With the spatial covariant derivative, spatial dot product and spatial
cross product, and obvious definitions of 
$ {\rm grad}_{\hbox{$u$}}$, 
$ {\rm curl}_{\hbox{$u$}}$, 
and $ {\rm div}_{\hbox{$u$}}$, 
one can mirror all the usual operations of 3-dimensional vector analysis,
or with the introduction of the spatial exterior derivative
$ {\rm d}(u) = P(u) {\rm d} $
and the spatial Lie bracket
$ [X,Y](u) = P(u)[X,Y]$,
all of the corresponding exterior derivative algebra. 

The various covariant derivatives of the spacetime and spatial metric
are all zero
\beq
    \nabla_{\rm(fw)}(u) g_{\alpha\beta}
    = \nabla_{\rm(fw)}(u) P(u)_{\alpha\beta} = 0\ ,\qquad
    \nabla(u)_\gamma g_{\alpha\beta}
    = \nabla(u)_\gamma P(u)_{\alpha\beta} = 0\ ,
\eeq
so index raising/lowering commutes with these derivatives.
The covariant derivative $\nabla_\beta u^\alpha$ may be measured
to yield the so called kinematical quantities of $u^\alpha$ 
\cite{ehl,haw,ell71,ell73}.
The measured fields are
a scalar (zero), a vector: the acceleration 
$a(u)^\alpha = [\nabla_{\rm(fw)}(u) u]^\alpha$,
a 1-form (zero), and a spatial tensor 
$\nabla(u)_\beta u^\alpha \equiv -k(u)^\alpha{}_\beta$,
the vanishing of the two fields due to the unit condition on $u^\alpha$:
$u^\alpha u_\alpha = -1$.

The spatial covariant derivative of $u^\alpha$ may in turn be decomposed
into its symmetric and antisymmetric parts, yielding
the vorticity (rotation) tensor $\omega(u)^\alpha{}_\beta$
and the expansion tensor $\theta(u)^\alpha{}_\beta$
(recall $[u^\flat]_\alpha = u_\alpha$)
\beq\eqalign{
   \omega(u)_{\alpha\beta} &= -\nabla(u)_{[\alpha} u_{\beta]}
                            = \half [d(u)u^\flat]_{\alpha\beta}  \ ,\\
   \theta(u)_{\alpha\beta} &= \nabla(u)_{(\alpha} u_{\beta)}
             = \nabla_{\rm(lie)}(u) g_{\alpha\beta}
             = \nabla_{\rm(lie)}(u) P(u)_{\alpha\beta}\ .}
\eeq
The spatial dual of the  vorticity tensor yields the spatial vorticity
(rotation) vector $[\vec\omega(u)]^\alpha =\omega(u)^\alpha$ 
(the ``overarrow" on the kernel symbol avoids ambiguity)
\beq
     \omega(u)^\alpha 
         = \half \eta(u)^{\alpha\beta\gamma}\omega(u)_{\beta\gamma} \ ,\qquad
    \omega(u)^\alpha{}_\beta X^\beta = - [\vec\omega(u) \times_u X]^\alpha\ .
\eeq
The expansion tensor may itself decomposed into its pure trace, the
expansion scalar 
$\Theta(u) = \theta(u)^\alpha{}_\beta = \nabla_\alpha u^\alpha$,
and the tracefree part, the shear tensor 
$\sigma(u)^\alpha{}_\beta 
= \theta(u)^\alpha{}_\beta - \frac13 \Theta(u) \delta^\alpha{}_\beta $\ .

The expansion, shear, and rotation describe the relative motion of neighboring
test observers with respect to a set of Fermi-propagated spatial triad vectors
along each test observer world line, which is encoded in the relationship
between the two types of temporal derivatives
\beq
  \nabla_{\rm(fw)}(u) X^\alpha 
     =   \nabla_{\rm(lie)}(u) X^\alpha + [\vec\omega(u) \times_u X]^\alpha
        + \theta(u)^\alpha{}_\beta X^\beta \ .
\eeq
A vector field $Y^\alpha$ is called a ``connecting vector" if
$\pounds_{\hbox{$u$}} Y^\alpha = 0$, i.e., if it is invariant under dragging
along by the flow of $u^\alpha$. If $Y^\alpha$ is small compared
to the characteristic distances over which $u^\alpha$ itself varies,
it may be interpreted as connecting a point on a 
given observer worldline to a point on a neighboring one whose position 
is identified with the tip of $Y^\alpha$ in the tangent space.
The spatial projection $X^\alpha = [P(u)Y]^\alpha$ may be interpreted
as the spatial position vector of this neighboring
observer in the local rest space $LRS_u$,
i.e., the position of the neighboring test observer as seen by the first one.
It satisfies $\nabla_{\rm(lie)}(u) X^\alpha = 0$, which means that
compared to a set of orthonormal spatial frame vectors $\{ e_a^\alpha\}$
which are Fermi-Walker transported along the $u$ congruence:
$\nabla_{\rm(fw)}(u) e_a^\alpha = 0$,
the ``relative position vector" $X^\alpha$ of neighboring observers
undergoes a combined scaling, (volume-preserving)
deformation, and rotation of the
local rest space $LRS_u$ whose rates
are determined respectively by the expansion scalar, the shear tensor,
and the vorticity tensor \cite{ehl,haw,ell71,ell73}.

The index-free formula
$ d u^\flat (X,Y) = X u^\flat(Y) - Y u^\flat(X) - u^\flat([X,Y])$
applied to spatial vector fields $X$ and $Y$ (orthogonal to $u_\alpha$)
immediately reduces to
\beq
    2 \vec\omega(u) \cdot (X \times_u Y) = 2\omega(u)^\flat(X,Y)
      = d u^\flat(X,Y) = - u^\flat([X,Y]) \ .
\eeq
The measurement of the Lie bracket of two spatial vector fields then becomes
\beq
      [X,Y]^\alpha = [X,Y](u)^\alpha + 2\omega^\flat(u)(X,Y) u^\alpha \ .
\eeq
The spatial Lie bracket of two spatial vector fields describes the
``closure of their quadrilateral" (see box 9.2 of Misner, Thorne, and Wheeler
\cite{mtw} and Fig.~2)
projected into the observer local rest space,
while (twice) the vorticity tensor evaluated on them describes the failure of
the two paths from the origin 
to the open vertex to remain synchronized with respect to the
observer, equaling the change in observer proper time between the two events.
Looked instead as a closed loop around the quadrilateral 
plus the commutator closer,
the latter factor describes the synchronization defect (change in observer
proper time from the beginning to the end) of the spatially closed loop 
in the local rest space of the observer.
\typeout{Extra sentence added here:}%
This in turn is intimately related to the Sagnac effect for a null curve
which has the same projection to the observer space \cite{ash,post}.

%%%%%%%%%%%%%%%%%%%%%%%%%%%%%%%%%%%%%%%%%%%%%%%%%%%%%%%%%%%%%%%%%%%%%%%%%%%%%%%
%  figure0: new figure between 1 and 2
\begin{figure}
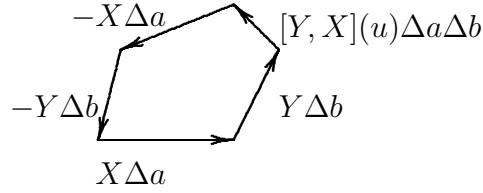


$$ \vbox{
\beginpicture
  \setcoordinatesystem units <.6cm,.6cm> point at 0 0 
%%%%%%%%%%%%%%%%%%%%%%%%%%%%%%%%%%%%%%%%%%%%%%%%%%%%%%%%%%%%%%%%%%%%%%%%%%%%%%%%
    \putrule from 0 0    to 3 0  
       
\setlinear
    \plot  3 0     4 2 /       
    \plot  4 2     3 3 /
    \plot  3 3     0.5 2 /      
    \plot  0.5 2   0 0 /       
\setsolid
%%%%%%%%%%%%%%%%%%%%%%%%%%%%%%%%%%%%%%%%%%%%%%%%%%%%%%%%%%%%%%%%%%%%%%%%%%%%%%%%
% text:a:
  \put {\mathput{X \Delta a}}                          [rt]   at  1.5 -0.5
  \put {\mathput{Y \Delta b}}                          [lt]   at  4.0  1.0
  \put {\mathput{[Y,X](u) \Delta a \Delta b}}          [lt]   at  4.0  2.8
  \put {\mathput{-X \Delta a}}                         [rt]   at  1.5  3.0
  \put {\mathput{-Y \Delta b}}                         [rt]   at  0.0  1.0
  
%%%%%%%%%%%%%%%%%%%%%%%%%%%%%%%%%%%%%%%%%%%%%%%%%%%%%%%%%%%%%%%%%%%%%%%%%%%%%%%%
% arrowheads:
\arrow <.3cm> [.1,.4]    from  2.6 0 to 3 0 
\arrow <.3cm> [.1,.4]    from 3.82 1.64 to 4 2
\arrow <.3cm> [.1,.4]    from 3.28 2.72 to 3 3
\arrow <.3cm> [.1,.4]    from .87 2.15  to .5 2
\arrow <.3cm> [.1,.4]    from .10 .39 to 0 0

%%%%%%%%%%%%%%%%%%%%%%%%%%%%%%%%%%%%%%%%%%%%%%%%%%%%%%%%%%%%%%%%%%%%%%%%%%%%%%%%
\endpicture}$$

\caption{
The commutator as a closer of quadilaterals. Measuring the ``closer"
commutator expression yields the spatial closer of the spatial projection
of the quadrilateral and the synchronization defect of the ``closed loop."
}\label{fig:0}
\end{figure}
%%%%%%%%%%%%%%%%%%%%%%%%%%%%%%%%%%%%%%%%%%%%%%%%%%%%%%%%%%%%%%%%%%%%%%%%%%%%%%%

\section{Measuring the Intrinsic Derivative Along a Curve}

Suppose one has an arbitrary parametrized curve $c(\lambda)$
in spacetime with tangent vector 
$c'(\lambda)^\alpha = [d c(\lambda)/ d\lambda]^\alpha$
and one wishes to measure the intrinsic or absolute derivative
$D/d\lambda$ along this curve.
This is important for measuring the equations of
motion of a test particle following a geodesic or under the
influence of some applied force or force field, or for studying
more general curves in spacetime. For example, one may easily
introduce a Serret-Frenet frame along any nonnull curve to study
the differential properties of the curve itself \cite{iyevis}
and then repeat the process from the point of view of the family
of test observers.

The intrinsic derivative along the curve is defined so that
if $X^\alpha$ is an arbitrary vector field on spacetime, then
(notationally suppressing the dependence of both sides on $c(\lambda)$)
\beq
       D X^\alpha / d \lambda = c'(\lambda)^\beta \nabla_\beta X^\alpha
             = d X^\alpha / d \lambda 
                + \Gamma^\alpha{}_{\beta\gamma} X^\gamma c'(\lambda)^\beta \ .
\eeq
It is then restricted to vector fields defined only along the curve so
that when extended smoothly to a vector field defined on the spacetime
the previous result is obtained, i.e., by using the second of the two
formulas.

Measuring the tangent vector
\beq
      c'(\lambda)^\alpha 
         \leftrightarrow ( - c'(\lambda)^\beta u_\beta, 
                                        [P(u) c'(\lambda)]^\alpha ) \ ,
\eeq
one can introduce two different new parametrizations of the curve
valid respectively when this tangent vector is not orthogonal to or
proportional to $u$ itself (not purely spatial or temporal)
\beq
        d\tau_{(c'(\lambda),u)} / d \lambda = - c'(\lambda)^\beta u_\beta \ ,
              \qquad
        d\ell_{(c'(\lambda),u)} / d \lambda = || P(u) c'(\lambda)||
\eeq
as illustrated in Fig.~3. %\ref{fig:3}.
The first re-parametrization corresponds to the (continuous) sequence of 
observer proper-time differentials of the 1-parameter family of observers
which cross paths with this curve, while the second corresponds to the
sequence of spatial arclength differentials of the relative motion
seen by this family.

%%%%%%%%%%%%%%%%%%%%%%%%%%%%%%%%%%%%%%%%%%%%%%%%%%%%%%%%%%%%%%%%%%%%%%%%%%%%%%%
% figure 2->3:
\begin{figure}
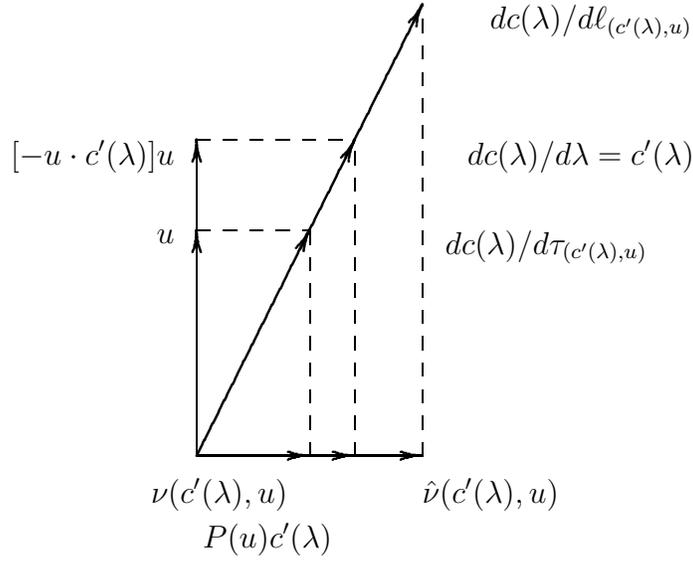


$$ \vbox{
\beginpicture
%  \twelvept
%  \setcoordinatesystem units <.8cm,.8cm> point at 0 100  %% to put baseline at top
  \setcoordinatesystem units <.6cm,.6cm> point at 0 0 
%%%%%%%%%%%%%%%%%%%%%%%%%%%%%%%%%%%%%%%%%%%%%%%%%%%%%%%%%%%%%%%%%%%%%%%%%%%%%%%%
    \putrule from 0 0   to 0 7         % ver leg   x u
    \putrule from 0 0   to 5 0         % hor leg   \nu(-,u)

    \plot 0 0  5.0  10.0  /                   % dc/ds

\setdashes
    \putrule from 3.5 0  to 3.5 7.0           % dc/d\lambda  vert
    \putrule from 0 7.0  to 3.5 7.0           % dc/d\lambda  hor

    \putrule from 2.5 0  to 2.5 5.0           % dc/d\tau vert
    \putrule from 0 5.0  to 2.5 5.0           % dc/d\tau  hor

    \putrule from 5.0 0  to 5.0 10.0          % dc/d\tau vert

\setsolid
%%%%%%%%%%%%%%%%%%%%%%%%%%%%%%%%%%%%%%%%%%%%%%%%%%%%%%%%%%%%%%%%%%%%%%%%%%%%%%%%
% text:a:
  \put {\mathput{u}}                          [rt]   at  -.5 5.0
  \put {\mathput{[-u\cdot c'(\lambda)] u}}  [rt]   at  -.5 7.0
  \put {\mathput{\nu(c'(\lambda),u)}}        [rt]   at  2.0 -0.5
  \put {\mathput{P(u) c'(\lambda)}}  [rt]   at  3.0 -1.5 
  \put {\mathput{ \hat\nu(c'(\lambda),u)}}    [lt]   at  5.0 -0.5
  \put {\mathput{dc(\lambda)/d\tau_{(c'(\lambda),u)}}} [lt]   at  5.5 5.0
  \put {\mathput{dc(\lambda)/d\lambda =c'(\lambda)}}  [lt]   at  6.0 7.0
  \put {\mathput{dc(\lambda)/d\ell_{(c'(\lambda),u)}}} [lt]   at  6.5 10.0

%%%%%%%%%%%%%%%%%%%%%%%%%%%%%%%%%%%%%%%%%%%%%%%%%%%%%%%%%%%%%%%%%%%%%%%%%%%%%%%%
% arrowheads:
% vertical:
    \arrow <.3cm> [.1,.4]    from  0 4.6 to 0 5.0            % u:
    \arrow <.3cm> [.1,.4]    from  0 6.6 to 0 7.0            % gamma'^{-1} u:
% slope 2/1:
    \arrow <.3cm> [.1,.4]    from  2.3 4.6 to 2.5 5.0      % dc/d\tau:
    \arrow <.3cm> [.1,.4]    from  3.3 6.6 to 3.5 7.0      % dc/d\lambda:
    \arrow <.3cm> [.1,.4]    from  4.8 9.6 to 5.0 10.0     % dc/ds:
% horizontal:
    \arrow <.3cm> [.1,.4]    from  2.1 0 to 2.5 0            % \nu(c,u):
    \arrow <.3cm> [.1,.4]    from  3.1 0 to 3.5 0            % P(u)c:
    \arrow <.3cm> [.1,.4]    from  4.6 0 to  5.0 0           % \hat\nu(c,u)

%%%%%%%%%%%%%%%%%%%%%%%%%%%%%%%%%%%%%%%%%%%%%%%%%%%%%%%%%%%%%%%%%%%%%%%%%%%%%%%%
\endpicture}$$

\caption{Measuring the tangent vector to a curve and re-parametrizing the
curve in terms of an observer-proper-time or observer-spatial-arclength
parameter.}\label{fig:3}
\end{figure}
%%%%%%%%%%%%%%%%%%%%%%%%%%%%%%%%%%%%%%%%%%%%%%%%%%%%%%%%%%%%%%%%%%%%%%%%%%%%%%%

One may introduce the relative velocity
\beq
      \nu(c'(\lambda),u)^\alpha 
            = [-c'(\lambda)^\beta u_\beta ]^{-1} [ P(u) c'(\lambda) ]^\alpha
\eeq
and the unit vector 
$\hat\nu(c'(\lambda),u)^\alpha
= ||\nu(c'(\lambda),u)||^{-1} \nu(c'(\lambda),u)^\alpha$
specifying the direction of relative motion
as long as the tangent vector is not spatial.
%(in which case one can
%normalize the original tangent vector to define this unit vector)
%
Then the two new parameters are related to each other by
\beq
      d \ell_{(c'(\lambda),u)} / d \tau_{(c'(\lambda),u)}
            = ||\nu(c'(\lambda),u)||
\eeq
as long as the tangent is not purely temporal or purely spatial.
Using the chain rule one can re-parametrize the intrinsic derivative
to correspond to these two new parametrizations
to obtain two new derivatives
$   D X^\alpha / d\tau_{(c'(\lambda),u)} $ and
$   D X^\alpha / d\ell_{(c'(\lambda),u)} $, for example
\beq
   D X^\alpha / d\tau_{(c'(\lambda),u)} 
       = [d\tau_{(c'(\lambda),u)} / d\lambda]^{-1} D X^\alpha / d\lambda \ .
%     \qquad
%    D X^\alpha / d\ell_{(c'(\lambda),u)} 
%       &= [d\ell_{(c'(\lambda),u)} / d\lambda]^{-1} D X^\alpha / d\lambda \ .}
\eeq

In order to measure the intrinsic derivative (for any of these 
parametrizations), one must introduce the Fermi-Walker
spatial intrinsic derivative along
the curve (which preserves the spatiality)
by simply taking its spatial projection,
leading to the spatial Fermi-Walker derivative 
$\nabla_{\rm(fw)}(u)$ (a temporal derivative)
acting along the temporal projection of the tangent 
and the spatial covariant derivative $\nabla(u)$ (a spatial derivative)
acting along the spatial projection of the tangent 
\beq
    D_{\rm(fw)}(c'(\lambda),u) / d\lambda 
        \equiv P(u) D / d\lambda 
        = [- c'(\lambda)^\beta u_\beta] \nabla_{\rm(fw)}(u)
            + [ P(u)c'(\lambda) ]^\beta  \nabla(u)_\beta \ .
\eeq
This is then reinterpreted as above as an operator for tensors defined
only along the parametrized curve \cite{mfg}.
The action of this operator on the family of spatial fields resulting from
the measurement of a tensor field along the curve together with 
kinematical linear transformations then leads to the family of spatial fields
which result from the measurement of the intrinsic derivative of the tensor
field.
For example, acting on $u^\alpha$ itself leads to a linear transformation
acting on the tangent vector
\beq
    D_{\rm(fw)}(c'(\lambda),u) u^\alpha / d\lambda 
        = [- c'(\lambda)^\beta u_\beta] a(u)^\alpha
            + k(u)^\alpha{}_\beta [ P(u)c'(\lambda) ]^\beta \ .
\eeq

How can one interpret the Fermi-Walker spatial intrinsic derivative?
Any derivative along a curve may be understood in terms of the associated
transport of a vector along it by requiring that the vector have zero
derivative along the curve: 
$D_{\rm(fw)}(c'(\lambda),u) X^\alpha / d \lambda = 0$.
Transporting a spatial vector in this way along the curve may be
thought of as the limit of a sequence of alternating steps along $u$
and orthogonal to $u$. Along $u$ it is Fermi-Walker transported
(unchanging with respect to a Fermi-Walker transported spatial frame),
while orthogonal to $u$ it is transported so that it appears to
move parallel to itself as seen by each observer whose path is crossed,
always remaining spatial. The phenomenon of spatial curvature is
illustrated by a small closed loop in the observer-space, namely a curve which
starts and ends on the same observer world line. Compared to a
vector which is Fermi-Walker transported along $u$ itself between these
two points (remaining at rest with respect to the observer family), 
the vector transported in this way along
the curve will undergo a slight rotation, which involves the action of the
Fermi-Walker spatial curvature tensor when the two points are close together.
%,
%as well as of the vorticity tensor, which describes effects related to the
%nonintegrability of the local rest spaces of the observer congruence.

It is also quite useful to introduce two new derivatives
along the curve
in which the Fermi-Walker transport along the observer congruence is
replaced by spatial Lie transport as in the interpretation of the
kinematical quantities themselves, where the corresponding transported frame
attempts to follow the nearby observers,
or alternatively
by a ``co-rotating Fermi-Walker transport" in which the corresponding spatial
transported frame only undergoes the additional rotation of those nearby
observers (compared to a Fermi-Walker transported spatial frame) 
without the expansion and shear of the spatially Lie dragged spatial
frame vectors. The latter transport, like Fermi-Walker transport,
preserves inner products, while the spatial Lie transport does not unless
the expansion tensor vanishes implying the observer congruence
consists of Killing vector trajectories, in which case it agrees with the
spatial co-rotating Fermi-Walker transport.

For a spatial vector field these three spatial intrinsic derivatives
differ in the same way as the corresponding temporal derivatives
\beq\eqalign{
  \nabla_{\rm(fw)}(u) X^\alpha 
     &=   \nabla_{\rm(lie)}(u) X^\alpha + [\vec\omega(u) \times_u X]^\alpha
        + \theta(u)^\alpha{}_\beta X^\beta \ ,\\  \nabla_{\rm(fw)}(u) X^\alpha 
     &=   \nabla_{\rm(cfw)}(u) X^\alpha + [\vec\omega(u) \times_u X]^\alpha\ ,}
\eeq
namely
\beq
    D_{\rm(tem)}(c'(\lambda),u) / d\lambda 
        = [- c'(\lambda)^\beta u_\beta] \nabla_{\rm(tem)}(u)
            + [ P(u)c'(\lambda) ]^\beta  \nabla(u)_\beta \ ,
       \qquad {\scriptstyle\rm tem \,=\, fw,cfw,lie} \ .
\eeq

By introducing the curvature tensors for each of these three kinds of 
spatial transport
\beq\eqalign{
 & \big\{ [ \nabla(u)\sub{X}, \nabla(u)\sub{Y}] -\nabla(u)\sub{[X,Y]}\big\}
           Z^\alpha \\
   &\quad= R_{\rm(tem)}(u)^\alpha{}_{\beta\gamma\delta}
                                          X^\gamma Y^\delta Z^\beta
        + 2\omega(u)_{\gamma\delta} X^\gamma Y^\delta  
                                     \nabla_{\rm(tem)}(u) Z^\alpha\ ,
    \quad  {\scriptstyle\rm tem \,=\, fw, cfw, lie } } \ ,
\eeq
one can extend the usual discussion of section 11.4 of Misner, Thorne,
and Wheeler \cite{mtw} of the result of parallel transport around
a small closed loop to the corresponding discussion of these transports
of spatial vectors around a small closed loop in observer-space corresponding
to a spacetime curve with a spatial tangent vector, leading to a short 
lapse of observer proper time (synchronization defect)
between the initial and final point on the observer worldline 
due to the nonintegrability of the observer local rest
spaces (nonvanishing vorticity) as illustrated in Fig.~2.
Transporting a spatial vector back along the observer worldline
to the original spacetime point of departure 
contributes an additional term (depending on the type of transport used)
to the relationship between the
second covariant derivative expression and
the spatial curvature of each type, 
so that the resulting spatial curvature tensor describes
the change in the transported vector compared to the original vector.

None of these three spatial curvature tensors in general has all of
the usual symmetry properties of an ordinary curvature tensor, but one
can introduce a fourth
``symmetry-obeying" spatial curvature 
tensor \cite{mfg,ferr65} by defining
\beq
 R_{\rm(sym)}(u)^{\alpha\beta}{}_{\gamma\delta} 
    = R_{\rm(cfw)}(u)^{[\alpha\beta]}{}_{\gamma\delta}
   -4 \theta(u)^{[\alpha}{}_{[\gamma} \omega(u)^{\beta]}{}_{\delta]}
\eeq
which does, and one may define its symmetric Ricci tensor 
$R_{\rm(sym)}(u)^\alpha{}_\beta$
and symmetric Einstein tensor $G_{\rm(sym)}(u)^\alpha{}_\beta$ 
by the usual formulas. 
The spatial curvature tensors and their contractions appear in the
measurement of the spacetime curvature tensors (Weyl and Riemann), 
the Ricci and Einstein tensors, and the curvature scalar.
In the case of zero vorticity
when the local rest spaces integrate to a family of spacelike hypersurfaces,
all four coincide with the Riemann curvature tensor of the induced
Riemannian metric on these hypersurfaces thought of as a spatial tensor;
for a stationary spacetime (zero expansion tensor), the Lie, corotating
Fermi-Walker and symmetry-obeying spatial curvatures agree and correspond
to the curvature tensor of the quotient space Riemannian metric.
These two extremes characterize
exactly the dominance of the slicing point of view
in the study of dynamical spacetimes and the exclusive use of the
threading point of view in the study of stationary spacetimes in the
context of exact solutions.

\section{Measuring the Equation of Motion of a Test Particle: The 
Electromagnetic Analogy}

Suppose $c(\tau)$ 
(set $\lambda=\tau$ above)
is the proper-time-parametrized world line of a test particle of nonzero mass 
$m$
with 4-velocity  $U^\alpha = [d c(\tau)/ d\tau]^\alpha$ 
and acceleration $a(U)^\alpha = DU^\alpha/ d\tau$.
(Only if $U^\alpha$ is itself a vector field on spacetime does
$DU^\alpha/ d\tau = \nabla_{\hbox{$U$}} U^\alpha$.)
If it is moving in spacetime under the influence of a force $f(U)^\alpha$,
which measures to $(\gamma(U,u){\cal P}(U,u),\gamma(U,u) F(U,u)^\alpha)$,
where ${\cal P}(U,u)$ is the observed power and $F(U,u)^\alpha$ the observed
spatial force as in special relativity,
one can measure the equation of motion 
$m a(U)^\alpha = f(U,u)^\alpha$, leading to a scalar and spatial vector
equation.

For example,
if the test particle has electric charge $q$ and
the applied force is the Lorentz force 
$f(U)^\alpha = q f^\alpha{}_\beta U^\beta$
due to an electromagnetic field $f_{\alpha\beta}$,
then the spatial force is
\beq\eqalign{
F(U,u)^\alpha
       &=q  \{ E(u)^\alpha 
                          + B(u)^\alpha{}_\beta \nu(U,u)^\beta \} \\
       &=q  \{ E(u)^\alpha 
            + [\nu(U,u) \times_u \vec B(u)]^\alpha \ ,}
\eeq
in terms of the observed electric and magnetic vector fields.
The measured equation of motion is then
\beq\eqalign{
     D_{\rm(tem)}(U,u) p(U,u)^\alpha / d \tau_{(U,u)}
         &=  F{}^{\rm(G)}_{\rm(tem)}(U,u)^\alpha + F(U,u)^\alpha\ ,\\
     D_{\rm(tem)}(U,u) E(U,u) / d \tau_{(U,u)}
         &= [F{}^{\rm(G)}_{\rm(tem)}(U,u)_\beta + F(U,u)_\beta]\,\nu(U,u)^\beta
                    \ ,}
\eeq
where, for example, the Fermi-Walker apparent spatial gravitational force is
\beq\eqalign{\label{eq:eqofmotion}
F{}^{\rm(G)}_{\rm(fw)}(U,u)^\alpha
       &=m \gamma(U,u) \{ -a(u)^\alpha 
                                  + k(u)^\alpha{}_\beta \nu(U,u)^\beta \}  \\
       &=m \gamma(U,u) \{ g(u)^\alpha 
                    + H_{\rm(fw)}(u)^\alpha{}_\beta \nu(U,u)^\beta \} \\
       &=m \gamma(U,u) \{ g(u)^\alpha 
            + \half [\nu(U,u) \times_u \vec H(u)]^\alpha{} 
            + H^{\rm(sym)}_{\rm(fw)}(u)^\alpha{}_\beta \nu(U,u)^\beta \} \ .}
\eeq
The clear analogy with electromagnetism leads $g(u)=-a(u)$ to be called
the gravitoelectric (GE) field, $ H(u)^\alpha = 2\omega(u)^\alpha$ the
gravitomagnetic (GM) vector field (its coefficient $1/2$ above
has the value 1 for the other choices),
and $ H_{\rm(tem)}(u)^\alpha{}_\beta$
the gravitomagnetic tensor field, whose symmetric part is proportional
to the expansion tensor of $u$ which in turn is proportional to the
spatial Lie derivative of either the spacetime or spatial metric.
(The gravitoelectromagnetic terminology is due to Thorne \cite{tho,thoprimac}.)
Thus the additional feature of spatial geometry in gravitation contributes
its effect on the left hand side
in the spatial intrinsic derivative through its spatial derivatives 
as well as on the right hand side
through its Lie (temporal) derivative in the symmetric
part of the GM tensor field (and of course through spatial inner products).

The electromagnetic analogy extends to the measurement of the Einstein
equations (together with the identity $d^2 u^\flat = 0$), which
leads to four equations (two pair of scalar and spatial
vector equations) which are a nonlinear
generalization for the gravito-vector fields
of equations analogous to the corresponding four
measured Maxwell equations for the electric and magnetic vector fields
and an spatial symmetric tensor equation
due to the additional feature of spatial geometry \cite{mfg}.
The linearized version
of these equations leads to equations rather similar to 
Maxwell's equations \cite{for,mfg,thomax,damsofxu}

\section{Measured Potentials for the Gravito-Fields}

The test observer covariant 4-velocity $u_\alpha$ 
and the spatial metric $P(u)_{\alpha\beta}$ act as potentials for the 
spatial gravitational force terms and spatial connection which appear
in the equation of motion of a test particle.
Although the GE and GM vector fields result from the measurement of
the exterior derivative of the 4-potential $[d u^\flat]_{\alpha\beta}$,
in order to have a scalar and spatial vector potential for the
gravito-vector fields analogous to the electromagnetic case, 
one must introduce a full splitting of spacetime, namely a parametrized
nonlinear reference frame consisting of a parametrized slicing and a threading,
with the appropriate causality properties for each point of view.
A parametrization for the threading (a spatial coordinate system) in addition
provides explicit potentials for the spatial connection coefficients themselves.

Suppose $\{t,x^a\}$ are local coordinates adapted to the parametrized
nonlinear reference frame. These provide an explicit representation of
the associated threading and/or slicing points of view.
One can then parametrize the spacetime metric components in this
local coordinate system in terms of the observer orthogonal decomposition
of the tangent space, introducing the lapse functions, shift vector field
and 1-form, and spatial metrics
\beq\eqalign{
\hbox{threading:\qquad} 
     & ds^2 = -M^2 (dt- M_a dx^a)^2 + \gamma_{ab} dx^a dx^b \ ,\\
\hbox{slicing:\qquad} 
     & ds^2 = -N^2 dt^2 + g_{ab} (dx^a + N^a dt) (dx^b + N^b dt) \ .
}\eeq

Fig.~4 compares the geometrical interpretation of the lapse and shift
in the threading and slicing points of view, using some index-free notation
(no indices but
$X^\flat$ for $X_\alpha$ or $\vec X$ for $X^\alpha$ when ambiguity exists).
In analogy with electromagnetism, these quantities act as the scalar
and vector potentials respectively for the gravito-vector fields
\beq\eqalign{
\hbox{threading:\qquad} 
     & g(m)_\alpha =  - \nabla(m)_\alpha \ln M 
                 -\nabla_{\rm(lie)}(m)_{\hbox{$\partial/\partial t$}}
                                                          \, M_\alpha \ ,\\
     & H(m)^\alpha = M [\curl_m \vec M]^\alpha\ ,\\
\hbox{slicing:\qquad} 
     & g(n)_\alpha =  - \nabla(n)_\alpha \ln N \ ,\\
     & H(n,\partial/\partial t)^\alpha 
                   = N^{-1} [\curl_n \vec N]^\alpha\ .
}\eeq
Note the absense of a vector potential term in the slicing GE
field and keep in mind from Fig.~4 that the relative velocities between the
test observer and the orthogonal-slicing/threading directions are
respectively $M M^\alpha$ and $N^{-1} N^\alpha$ when comparing the 
two GM vector relationships.
One can also express the gravitomagnetic tensor field and the 
spatial connection coefficients in terms of the measured metric quantities.

The slicing GM vector field still must be defined, and differs from
the vanishing GM vector field of the corresponding hypersurface point of view
(zero vorticity) due to the new choice of evolution direction along

\typeout{figure 4}
\begin{figure}[ht]

%%%%%%%%%%%%%%%%%%%%%%%%%%%%%%%%%%%%%%%%%%%%%%%%%%%%%%%%%%%
\def\ds{} %\displaystyle
\def\mathput#1{\relax \ifmmode \ds #1\else $\ds #1$\fi}
\def\mathputvc#1{\mathput{\vcenter{\hbox{$\ds #1$}}}}
\def\leftbraceto#1{\mathputvc{\,\left\{
   \vcenter{\hbox{\vrule height #1 depth 0pt width 0pt}}
      \right.\,}}
\def\rightbraceto#1{\mathputvc{\,\left.
   \vcenter{\hbox{\vrule height #1 depth 0pt width 0pt}}
     \right\}\,}}

$$ \vbox{
\beginpicture
%  \twelvept

%  \setcoordinatesystem units <1cm,1cm> point at 0 0  
  \setcoordinatesystem units <.8cm,.8cm> point at 0 0  
%%%%%%%%%%%%%%%%%%%%%%%%%%%%%%%%%%%%%%%%%%%%%%%%%%%%%%%%%%%%%%%%%%%%%%%%%%%%%%%%
  \put {\mathput{Mm=\partial/\partial t}} [lt]   at  1.5    2.5
  \put {\mathput{m}}                      [lt]   at  2.3    4.9
  \put {\mathput{-M\vec M}}               [rB]   at  1.0    5.0
  \put {\mathput{-\vec M}}                [rt]   at  -0.5   2.0

  \put {\mathput{dt = -M^{-1} m^\flat +M^\flat}}   [lB]    at  4.5 3.5
  \put {\mathput{M^\flat}}                [l]    at  10.0   4.5
  \put {\mathput{-M^{-1} m^\flat}}        [rB]   at  5.5   4.8
  \put {\mathput{LRS_m}}                  [l]   at 10.5   3.5

%  \put {\mathput{\hbox{THREADING}}}       [B] at 4.0 -1.5
  \put {\mathput{\bullet}}                      at 0 0

  \put {\mathput{T(\hbox{slicing})}}       [rt]   at 10.0 -0.3

%%%%%%%%%%%%%%%%%%%%%%%%%%%%%%%%%%%%%%%%%%%%%%%%%%%%%%%%%%%%%%%%%%%%%%%%%%%%%%%%
  \setsolid
    \putrule from  0   0  to  0     4.36     % perp to dt planes
    \putrule from  -1  0  to  10.0  0        % dt axis
    \putrule from  -1  3  to  10.0  3        % dt upper
    \putrule from  -0.3  0  to  -0.3  0.3        % perp up
    \putrule from  -0.3  0.3  to  0   0.3        % perp across

%%%%%%%%%%%%%%%%%%%%%%%%%%%%%%%%%%%%%%%%%%%%%%%%%%%%%%%%%%%%%%%%%%%%%%%%%%%%%%%%
  \setlinear
    \plot  -1  -.33   10  3.33 /             % -M^{-1}m^\flat lower
    \plot  -1  2.33   11  6.33 /             % -M^{-1}m^\flat upper
    \plot  0  4.36    1.63 4.9 /             % -M^{-1}\vec M
    \plot  -.33  -1.0 2.0  6   /             %  M^\flat lower
    \plot  7.67  -1.0 10.33 7  /             %  M^\flat upper

    \plot  0.1 0.3    0.4 0.4 /             %  perp across
    \plot  0.3 0.1    0.4 0.4 /             %  perp up

%%%%%%%%%%%%%%%%%%%%%%%%%%%%%%%%%%%%%%%%%%%%%%%%%%%%%%%%%%%%%%%%%%%%%%%%%%%%%%%%
    \arrow <.3cm> [.1,.4]    from  0.9 2.7  to  1.0  3.0      % Mm
    \arrow <.3cm> [.1,.4]    from  1.53 4.6  to  1.63  4.9    % m

    \arrow <.3cm> [.1,.4]    from  0.3 2.77  to  0 2.67       % -\vec M
    \arrow <.3cm> [.1,.4]    from  0.3 4.46  to  0 4.36       % -\vec M

%%%%%%%%%%%%%%%%%%%%%%%%%%%%%%%%%%%%%%%%%%%%%%%%%%%%%%%%%%%%%%%%%%%%%%%%%%%%%%%%
% b):

%  \setcoordinatesystem units <1cm,1cm> point at -13.5 0  
  \setcoordinatesystem units <.8cm,.8cm> point at -13.5 0  

%%%%%%%%%%%%%%%%%%%%%%%%%%%%%%%%%%%%%%%%%%%%%%%%%%%%%%%%%%%%%%%%%%%%%%%%%%%%%%%%
  \put {\mathput{Nn}}                     [rt]   at  -0.5    3.0
  \put {\mathput{n}}                      [rt]   at  -0.5    4.5
  \put {\mathput{N^{-1}\vec N}}           [B]   at  0.7    5.0
  \put {\mathput{\vec N}}                 [B]   at   0.5    3.5

  \put {\mathput{\partial/\partial t = Nn+\vec N}} [lt]   at  1.5   3.0
  \put {\mathput{N^{-1} \partial/\partial t}} [lt]   at  2.0   4.5
  \put {\mathput{LRS_n}}                  [l]   at 2.5   0

%  \put {\mathput{\hbox{SLICING}}}          [B]     at 0.5 -1.5
  \put {\mathput{\bullet}}                      at 0 0
  \put {\mathput{T(\hbox{threading})}}           [B]   at 2 6.5

%%%%%%%%%%%%%%%%%%%%%%%%%%%%%%%%%%%%%%%%%%%%%%%%%%%%%%%%%%%%%%%%%%%%%%%%%%%%%%%%
  \setsolid
    \putrule from  0   0  to  0     4.5      % perp to dt planes
    \putrule from  -1  0  to  2     0        % LRS_n hor axis
    \putrule from   0  3  to  1  3           % \vec N
    \putrule from   0  4.5  to  1.5  4.5     % N^{-1}\vec N
    \putrule from  -0.3  0  to  -0.3  0.3        % perp up
    \putrule from  -0.3  0.3  to  0   0.3        % perp across

%%%%%%%%%%%%%%%%%%%%%%%%%%%%%%%%%%%%%%%%%%%%%%%%%%%%%%%%%%%%%%%%%%%%%%%%%%%%%%%%
  \setlinear
    \plot  0 0 2 6 /                    % threading tangent N^{-1} d/dt

%%%%%%%%%%%%%%%%%%%%%%%%%%%%%%%%%%%%%%%%%%%%%%%%%%%%%%%%%%%%%%%%%%%%%%%%%%%%%%%%

    \arrow <.3cm> [.1,.4]    from  0.9 2.7  to  1.0  3.0      % d/dt
    \arrow <.3cm> [.1,.4]    from  1.4 4.2  to  1.5  4.5     % N^{-1} d/dt

    \arrow <.3cm> [.1,.4]    from  0 4.2    to  0 4.5       %  n
    \arrow <.3cm> [.1,.4]    from  0 2.7    to  0 3         % N n

    \arrow <.3cm> [.1,.4]    from  1.2 4.5    to  1.5 4.5       % N^{-1}\vec N
    \arrow <.3cm> [.1,.4]    from  0.7 3.0    to  1.0   3       %  \vec N

%%%%%%%%%%%%%%%%%%%%%%%%%%%%%%%%%%%%%%%%%%%%%%%%%%%%%%%%%%%%%%%%%%%%%%%%%%%%%%%%
% c):
%  \setcoordinatesystem units <1cm,1cm> point at 0 6
  \setcoordinatesystem units <.8cm,.8cm> point at 0 6
%%%%%%%%%%%%%%%%%%%%%%%%%%%%%%%%%%%%%%%%%%%%%%%%%%%%%%%%%%%%%%%%%%%%%%%%%%%%%%%%
  \put {\mathput{\delta t \,\partial/\partial t}} [rB]   at  0 1.5

  \put {\mathput{x^i}}                      [lt]   at  1.7    4
  \put {\mathput{x^i + \delta x^i}}         [rt]   at  8.8    4

  \put {\mathput{t}}                      [r]   at  -1.5    0
  \put {\mathput{t + \delta t}}           [r]   at  -1.5   3

  \put {\mathput{LRS_m}}                  [lB]   at 9.7   3.8

%  \put {\mathput{\hbox{THREADING}}}       [B] at 4.0 -1.3
  \put {\mathput{\hbox{THREADING}}}       [B] at 4.0 -1.1

  \put {\mathput{\delta t = M_i \delta x^i\ ,\ \ 
                  \delta\tau_m = M \delta t}}    [t]   at 4   -1.3
%                  \delta\tau_m = M \delta t}}    [t]   at 4   -1.5

%%%%%%%%%%%%%%%%%%%%%%%%%%%%%%%%%%%%%%%%%%%%%%%%%%%%%%%%%%%%%%%%%%%%%%%%%%%%%%%%
  \setsolid
    \putrule from  -1  0  to  10.0  0        % dt axis
    \putrule from  -1  3  to  10.0  3        % dt upper

%%%%%%%%%%%%%%%%%%%%%%%%%%%%%%%%%%%%%%%%%%%%%%%%%%%%%%%%%%%%%%%%%%%%%%%%%%%%%%%%
  \setlinear
    \plot  -1  -.33   10  3.33 /             % -M^{-1}m^\flat lower
    \plot  -.33  -1.0 1.33  4   /             %  M^\flat lower
    \plot  7.67  -1.0 9.33  4  /             %  M^\flat upper

    \plot  0.1 0.3    0.4 0.4 /             %  perp across
    \plot  0.3 0.1    0.4 0.4 /             %  perp up

%%%%%%%%%%%%%%%%%%%%%%%%%%%%%%%%%%%%%%%%%%%%%%%%%%%%%%%%%%%%%%%%%%%%%%%%%%%%%%%%
    \arrow <.3cm> [.1,.4]    from  0.9 2.7  to  1.0  3.0      % Mm

%%%%%%%%%%%%%%%%%%%%%%%%%%%%%%%%%%%%%%%%%%%%%%%%%%%%%%%%%%%%%%%%%%%%%%%%%%%%%%%%
% d):
%  \setcoordinatesystem units <1cm,1cm> point at -13.5 6
  \setcoordinatesystem units <.8cm,.8cm> point at -13.5 6

%%%%%%%%%%%%%%%%%%%%%%%%%%%%%%%%%%%%%%%%%%%%%%%%%%%%%%%%%%%%%%%%%%%%%%%%%%%%%%%%
  \put {\mathput{x^i}}                    [lt]   at  1.8   4
  \put {\mathput{x^i - \delta x^i}}       [rt]   at  -0.2  4

  \put {\mathput{t}}                      [r]   at  -1.5    0
  \put {\mathput{t+\delta t}}                      [r]   at  -1.5    3.0

  \put {\mathput{\delta t \,\partial/\partial t}} [lB]   at  1.0 1.5

%  \put {\mathput{\hbox{SLICING}}}          [B]     at 0.5 -1.3
  \put {\mathput{\hbox{SLICING}}}          [B]     at 0.5 -1.1

  \put {\mathput{\delta x^i = N^i \delta t\ ,\ \ 
              \delta\tau_n = N \delta t}}    [t]   at 0.5   -1.3
%              \delta\tau_n = N \delta t}}    [t]   at 0.5   -1.5

  \put {\mathput{LRS_n}}                  [l]   at 2.5   0

%%%%%%%%%%%%%%%%%%%%%%%%%%%%%%%%%%%%%%%%%%%%%%%%%%%%%%%%%%%%%%%%%%%%%%%%%%%%%%%%
  \setsolid
    \putrule from  0   0  to  0     3      % perp to dt planes
    \putrule from  -1  0  to  2     0        % t
    \putrule from  -1  3  to  2     3        % t = dt
    \putrule from  -0.3  0  to  -0.3  0.3        % perp up
    \putrule from  -0.3  0.3  to  0   0.3        % perp across

%%%%%%%%%%%%%%%%%%%%%%%%%%%%%%%%%%%%%%%%%%%%%%%%%%%%%%%%%%%%%%%%%%%%%%%%%%%%%%%%
  \setlinear
    \plot   0 0 1.33 4 /                    % x^i
    \plot  -1 0 .33 4  /                    % x^i - dx^i

%%%%%%%%%%%%%%%%%%%%%%%%%%%%%%%%%%%%%%%%%%%%%%%%%%%%%%%%%%%%%%%%%%%%%%%%%%%%%%%%

    \arrow <.3cm> [.1,.4]    from  0.9 2.7  to  1.0  3.0      % d/dt

%%%%%%%%%%%%%%%%%%%%%%%%%%%%%%%%%%%%%%%%%%%%%%%%%%%%%%%%%%%%%%%%%%%%%%%%%%%%%%%%

%%%%%%%%%%%%%%%%%%%%%%%%%%%%%%%%%%%%%%%%%%%%%%%%%%%%%%%%%%%%%%%%%%%%%%%%%%%%%%%%

\endpicture}$$
\vfill
\centerline{Figure~3. See adjacent page for caption.} 

%%%%%%%%%%%%%%%%%%%%%%%%%%%%%%%%%%%%%%%%%%%%%%%%%%%%%%%%%%%
\end{figure}

%%%%%%%%%%%%%%%%%%%%%%%%%%%%%%%%%%%%%%%%%%%%%%%%%%%%%%%%%%%%%%%%%%%%%%%%%%%%%%%
% figure 3_4:
\typeout{Figure 4:}
\begin{figure}[ht]
%\input fig$tex:mg7bobfig3      % caption only; no room on full page figure
%\vglue6in
\caption{%
%$(a)$
\null\hfill\break
{\bf Top:}
Tangent space relationship of observer splitting to the parametrized
nonlinear reference frame splitting.
In each case the tangent space  horizontal axis is tangent to the
slicing and the vertical axis is orthogonal to it (nonnull slice!).
Each of the three pairs of parallel lines in the threading diagram
is a cross-section of a pair of parallel 3-planes in the tangent space
representing the corresponding 1-form as described by 
Burke \protect\cite{bur},  % 14
with the three pairs related to each other by 1-form addition
describing the relationship of the differential $dt$ to the 
threading observer. This is analogous to the vector relationship
between $\partial/\partial t$ and the slicing observer 4-velocity.
Each of these
defines the lapse and shift in the corresponding point of view.\hfil\break
%
%$(b)$
{\bf Bottom:}
Infinitesimal displacement relationship of observer splitting to the
parametrized nonlinear reference frame. 
In the threading p.o.v.\ the shift 1-form describes the change in $t$
towards the observer local rest space as one moves along the slicing,
while in the slicing p.o.v.\ the shift vector field describes the
change in $x^i$ away from the observer local time direction as one advances 
along the threading. The lapse in each case converts the coordinate
time changes to observer proper time changes.
}% end caption
%\label{fig:4}
\end{figure}
%%%%%%%%%%%%%%%%%%%%%%%%%%%%%%%%%%%%%%%%%%%%%%%%%%%%%%%%%%%%%%%%%%%%%%%%%%%%%%%

\noindent
the distinct threading rather than along the normal congruence to the
slicing. This brings us to the question of evolution in the slicing
point of view, which is a hybrid or bi-congruence approach.

In this point of view
the measurement of tensor equations involving covariant derivatives
naturally leads to measured expressions involving the Fermi-Walker
temporal derivative 
which may be re-expressed in terms of the Lie temporal derivative
$\nabla_{\rm(fw)}(n) X^\alpha  
 = \nabla_{\rm(lie)}(n) X^\alpha + \theta(n)^\alpha{}_\beta X^\beta$
which in turn may be re-expressed in terms of
the threading temporal derivative
$\nabla_{\rm(lie)}(n) 
 = \nabla_{\rm(lie)}(n,\partial/\partial t)
             - N^{-1} \pounds(n)_{\hbox{$\vec N$}}$,
where
$ \nabla_{\rm(lie)}(n,\partial/\partial t)
  = N^{-1} P(n) \pounds(n)_{\hbox{$\partial/\partial t$}}$.
This spatially projected Lie derivative along the threading is the
evolution operator in the slicing point of view, and the slicing spatial
intrinsic derivative is defined by using
this temporal operator in its equivalent
decomposition when acting on spacetime tensor fields.

Given a test particle word line with 4-velocity 
\beq\eqalign{
     U^\alpha &= \gamma(U,n) \{ n^\alpha + \nu(U,n)^\alpha \} \\
              &= \gamma(U,n) \{ N^{-1} \partial/\partial t 
                                 + [\nu(U,n)^\alpha - N^{-1} N^\alpha] \}
     \ ,}
\eeq
the latter decomposition into threading and slicing parts carries over
to the slicing spatial intrinsic derivative
\beq
\eqalign{
 D_{\rm(lie)}(U,n,\partial/\partial t) X^\alpha / d\tau
    &= \gamma(U,n) [ \del_{\rm(lie)}(n,\partial/\partial t) X^\alpha
               + \del(n)\sub{[\nu(U,n) - N^{-1}\vec N]} X^\alpha ]\\
      &= D_{\rm(lie)}(U,n) X^\alpha / d\tau 
   -\gamma(U,n)\Delta H_{\rm(lie)}(n,\partial/\partial t)^\alpha{}_\beta 
         X^\beta ] \ ,
}
\eeq
where the first equality only holds for spacetime vector fields and
the difference term 
\beq\eqalign{
  \Delta H_{\rm(lie)}(n,\partial/\partial t)^\alpha{}_\beta X^\beta
      &= N^{-1} [ \nabla(n)_{\hbox{$\vec N$}} 
                    - \pounds(n)_{\hbox{$\vec N$}} ] X^\beta \ ,\\
  [\Delta H_{\rm(lie)}(n,\partial/\partial t)]^\alpha{}_\beta 
      &= N^{-1} [\nabla(n) \vec N ]^\alpha{}_\beta
      = N^{-1} \nabla(n)_\beta  N^\alpha \ ,\cr
}\eeq
must  be subtracted from the hypersurface Lie spatial gravitational force
to obtain the slicing version. Combined with the
expansion tensor term already present,
this leads to the gravitomagnetic tensor (covariant form)
\beq\eqalign{
      H_{\rm(lie)}(n,\partial/\partial t)_{\alpha\beta}
        &= - N^{-1} \nabla(n)_\beta N_\alpha - 2 \theta(n)_{\alpha\beta} \\
        &=  N^{-1} \nabla(n)_\alpha N_\beta -
   \nabla_{\rm(lie)}(n,\partial/\partial t)g_{\alpha\beta} \ ,}
\eeq
whose antisymmetric part contributes the slicing gravitomagnetic
vector field to the equation of motion analogous to 
Eq.~(\ref{eq:eqofmotion}) with a multiplicative factor
of $1/2$ like the Fermi-Walker such coefficient. Thus in the slicing
point of view, gravitomagnetic effects arise from the relative motion
of the observers compared to the threading curves.

%%%%%%%%%%%%%%%%%%%%%%%%%%%%%%%%%%%%%%%%%%%%%%%%%%%%%%%%%%%%%%%%%%%%%%%%%%%%%%%
%\newpage\typeout{New page:}

\section{Just a Big Word?}

Okay, I think I have given enough of a sketch of the key ideas of the
splitting game from all points of view for a general review of this kind.
More details may be found elsewhere, but I wanted to present here a 
flavor of the foundations one can uncover when one starts to dig into this
subject and tries to formulate a language that encompasses all that one finds.
We can now return to the question posed by the title of this contribution.
``Gravitoelectromagnetism"---obviously it's a long word, but does the body
of ideas behind it stand up on its own? 
Or, sure we can play these mathematical
games, but are they actually useful? Does this more careful way of looking at
the various possibilities help us understand any better the spacetime
structure of interesting spacetimes or of various approaches to solving real
problems in gravitational theory?

I would claim that there are many instances where these ideas do give
valuable insight into various aspects of general relativity.
Let me just mention a few more recent examples.

Black hole spacetimes provide a rich arena for the exploration of ideas
in general relativity. As stationary axially symmetric spacetimes, they
have both a preferred threading by a timelike Killing vector field
whose associated test observers are called the static observers
or the distantly nonrotating observers, and
a preferred slicing by a family of spacelike hypersurfaces orthogonal
to the so called ``Zero Angular Momentum Observers" (ZAMO's)
or locally nonrotating observers. 
One can thus study all of the various splittings in these
exact model spacetimes and use them to better understand their properties.

In this context one can better appreciate the mathematical
structure of the problem of describing the precession of a test gyro relative
to the fixed stars, without confusing the problem with the linearization
which occurs in the post-Newtonian approximation to general 
relativity, the context in which it is usually studied, 
as first explored by Schiff and revisited by many others.
The exact Schiff precession formula is easily obtained in this
setting \cite{mfg}.

Thorne et al\cite{thoprimac}
have capitalized on the slicing point of view approach
to exact and approximate black hole problems, but simultaneously trying
to appeal to an audience of nonrelativists (but see Ref.~\citen{thomacmnras}).
The more general setting
of GEM helps to better understand that work and extend it to more general
spacetimes and to the
language of other splittings which may also prove useful in aiding
understanding of this topic.

Damour, Soffel and Xu \cite{damsofxu}
have written a beautiful series of articles 
on the  post-Newtonian celestial mechanics of extended bodies
which crucially uses the GEM tools in its analysis without directly
acknowledging their roots in the fully nonlinear context of general
relativity. Some of these connections have been described \cite{mfg,bcjw}.
The threading gravitoelectric and gravitomagnetic
vector fields are crucial to their approach.

Recently both Ehlers and Ellis have led the way in a renewed wave of
applications of the congruence point of view related to
both the ``Newtonian limit" \cite{ehl89}
and perturbations of cosmological models
as well as some classes of exact (but not solvable) classes
of  cosmological spacetimes. The geometry of the
congruence point of view is extremely powerful in the problem of
gauge invariant perturbation theory for Friedmann-Robertson-Walker spacetimes,
which appears much simpler to interpret than the more traditional
slicing point of view work led by Bardeen \cite{bar}.
(For references of work by
Ellis, Bruni, and collaborators, see Refs.~\citen{ellbru,ell94,elldun}.)
The Newtonian limit is also best described in terms of a 1-parameter
family of spacetimes with a congruence/threading splitting, and the geometry
of the Newtonian limit arises from this splitting.

This is just a sampling of some recent work in which splitting plays a crucial
role. In most cases, no attempt has been made to relate 
individual results to the larger context of other splitting approaches.

%%%%%%%%%%%%%%%%%%%%%%%%%%%%%%%%%%%%%%%%%%%%%%%%%%%%%%%%%%%%%%%%%%%%%%%%%%%%%%%
%\newpage\typeout{New page:}

\section{Concluding Remarks}

Before closing this sketch of gravitoelectromagnetism,
I should caution you that we don't want to exaggerate the importance
of these mathematical tools.
They merely provide a framework in which to interpret or perform
calculations and understand relationships between alternative ways of
looking at certain problems. Solving these problems is another
question altogether.

In the same way that the slicing approach of ADM has facilitated
many results in widely diverse areas of relativity,
we think that extending this approach to include all the possibilities
increases the power of the related set of tools.
Although in and of themselves these tools are not fundamental,
they do help us better appreciate the larger picture into which
many more aspects of general relativity fit together.
Certainly they illuminate the space-plus-time and therefore
the spacetime structure of many interesting applications of general
relativity theory and as such are clearly deserving of more attention
in the relativity community.
% perhaps even worthy of a Marcel Grossmann prize as an institution of
%ideas rather than of people.

%%%%%%%%%%%%%%%%%%%%%%%%%%%%%%%%%%%%%%%%%%%%%%%%%%%%%%%%%%%%%%%%%%%%%%%%%%%%%%%
%\newpage\typeout{New page:}

%%%%%%%%%%%%%%%%%%%%%%%%%%%%%%%%%%%%%%%%%%%%%%%%%%%%%%%%%%%%%%%%%%%%%%%%%%%%
%%%%%%%%%%%%%%%%%%%%%%%%%%%%%%%%%%%%%%%%%%%%%%%%%%%%%%%%%%%%%%%%%%%%%%%%%%%%

\begin{thebibliography}{00}
\textheight=8.75in

\bibitem{for} 
R.L. Forward, {\it Proceedings of the IRE\/} {\bf 49} (1961) 892.

\bibitem{adm} 
R. Arnowit, S. Deser and C.W. Misner, in {\it Gravitation: An 
Introduction to Current Research\/}, ed.
L. Witten (Wiley, New York, 1962).

\bibitem{mtw} 
C.W. Misner, K.S. Thorne and J.A. Wheeler, 
{\it Gravitation\/} (Freeman, San Francisco, 1973).

\bibitem{whe}
J.A. Wheeler, in {\it Relativity, Groups, and Topology\/},
eds. C. DeWitt and B.S. DeWitt (Gordon and Breach, New York, 1964).

\bibitem{ll}
I.D. Landau and E.M. Lifshitz, 
{\it The Classical Theory of Fields\/}
(Permagon, Oxford, 1983).

\bibitem{ehl} 
J. Ehlers,  
{\it Akad.\ Wiss.\ Mainz Abh., Math.-Nat.\ Kl.\/},  
Nr. 11 (1961) 793
[English translation by G.F.R.\ Ellis, 
    {\it Gen.\ Relativ.\ Grav.\/} {\bf 25} (1993) 1225].

\bibitem{haw} 
S.W. Hawking, 
{\it Astrophys.\ J.\/} {\bf 145} (1966) 544.

\bibitem{ell71} 
G.F.R. Ellis, 
in {\it General Relativity and Cosmology: Proceedings
of Course 47 of the International School of Physics `Enrico Fermi'\/},
ed. R. Sachs (Academic Press, New York, 1971).

\bibitem{ell73} 
G.F.R. Ellis, 
in {\it C\`argese Lectures in Physics\/} Vol.\ 6,
ed. E. Schatzman (Gordon and Breach, New York, 1973).

\bibitem{mfg}
R.T. Jantzen, P. Carini, and D. Bini,
{\it Ann.\ Phys.\ NY\/}
{\bf 215} (1992) 1.

\bibitem{iyevis}
B.R. Iyer and C.V. Vishveshwara, 
{\it Phys.\ Rev.\/} {\bf D48} (1993) 5706.

\bibitem{ferr65} 
G. Ferrarese, 
{\it Rend.\ di Mat.} {\bf 22} (1963) 147;
{\bf 24} (1965) 57.

\bibitem{tho} 
K.S. Thorne, 
  in {\it Quantum Optics, Experimental Gravitation, and Measurement Theory\/}
  eds.\ P.\ Meystre and M.O.\ Scully
  (Plenum Press, New York and London, 1981);
  in {\it Near Zero: New Frontiers of Physics\/}
  eds.\ J.D. Fairbank, B.S. Deaver, Jr., C.W. Everitt, and P.F. Michelson
  (Freeman, New York, 1988).

\bibitem{thoprimac} 
K.S. Thorne, R.H.  Price, and D.A. Macdonald, eds.,
  {\it Black Holes: The Membrane Paradigm\/}
  (Yale University Press, New Haven, 1986).

\bibitem{thomax}
V.B. Braginsky, C.M. Caves, and K.S. Thorne, 
  {\it Phys.\ Rev.\/} {\bf 15} (1977) 2047.

\bibitem{damsofxu}
T. Damour, M. Soffel, and C. Xu,
  {\it Phys.\ Rev.\/} {\bf D43} (1991) 3273;
                      {\bf D45} (1992) 1017;
                      {\bf D47} (1993) 3124.

\bibitem{bur}
W.K. Burke,
  {\it Applied Differential Geometry\/}
  (Cambridge University Press, Cambridge, 1985).

\bibitem{thomacmnras}
K.S. Thorne and D. Macdonald, 1982,  
  {\it M.N.R.A.S.} {\bf 198}, 339.

\bibitem{bcjw}
D. Bini, P. Carini, R.T. Jantzen, and D. Wilkins,
  {\it Phys.\ Rev.\/} {\bf D49} (1994) 2820.

\bibitem{bar}
J.M. Bardeen,
  {\it Astrophys.\ J.\/}  {\bf 162} (1970) 71.

\bibitem{ellbru} 
G.F.R.\ Ellis and M.\ Bruni, {\it Phys.\ Rev.\/} {\bf D40} (1989) 1804.

\bibitem{ell94}
G.F.R. Ellis,
  {\it The Covariant and Gauge Invariant Approach to Perturbations
      in Cosmology\/}, preprint 1994/16 
  (Dept.\ of Applied Math., University of Capetown).
  
\bibitem{ehl89}
J. Ehlers,
  {\it The Newtonian Limit of General Relativity\/},
  preprint 1989
  (Dept. of Applied Math., University of Cape Town).

\bibitem{elldun}
G.F.R. Ellis and P.K.S. Dunsby, 
  {\it Newtonian Evolution of the Weyl Tensor\/},
  e-print: astro-ph/9410001.

\bibitem{post}
E.J. Post
  {\it Rev.\ Mod.\ Phys.\/} {\bf 39} (1967) 475.

\bibitem{ash}
A. Ashtekar and A. Magnon,
  {\it J.\ Math.\ Phys.\/} {\bf 16} (1975) 341.

\end{thebibliography}
\end{document}